\documentclass{article}
\usepackage{spconf,amsmath,graphicx}
\usepackage{booktabs}

\usepackage{kotex}
\usepackage{amsmath}
\usepackage{bm}
\usepackage{amssymb}
\usepackage{hhline}
\usepackage{amsthm}
\usepackage{amsfonts}
\usepackage{commath}
\usepackage{mathtools}
\usepackage{nicefrac}
\usepackage{xfrac}
\usepackage{multirow}
\usepackage{color}
\usepackage{nccmath}
\usepackage{hyperref}
\usepackage{float}

\usepackage{pifont}
\newcommand{\cmark}{\ding{51}}
\newcommand{\xmark}{\ding{55}}

\title{REAL-TIME DENOISING AND DEREVERBERATION WTIH TINY RECURRENT U-NET}
\name{Hyeong-Seok Choi$^{1,2}$, Sungjin Park$^{1}$, Jie Hwan Lee$^{2}$, Hoon Heo$^2$, Dongsuk Jeon$^{1}$, Kyogu Lee$^{1,2}$}
\address{
  $^1$Department of Intelligence and Information, Artificial Intelligence Institute, Seoul National University \\$^2$Supertone Inc.}

\begin{document}
\ninept
\maketitle
\begin{abstract}
Modern deep learning-based models have seen outstanding performance improvement with speech enhancement tasks.
The number of parameters of state-of-the-art models, however, is often too large to be deployed on devices for real-world applications.
To this end, we propose Tiny Recurrent U-Net (TRU-Net), a lightweight online inference model that matches the performance of current state-of-the-art models.
The size of the quantized version of TRU-Net is 362 kilobytes, which is small enough to be deployed on edge devices.
In addition, we combine the small-sized model with a new masking method called phase-aware $\beta$-sigmoid mask, which enables simultaneous denoising and dereverberation.
Results of both objective and subjective evaluations have shown that our model can achieve competitive performance with the current state-of-the-art models on benchmark datasets using fewer parameters by orders of magnitude.
\end{abstract}
\vspace{-0.1in}
\section{Introduction}
\vspace{-0.1in}
\label{sec:intro}
In this paper, we focus on developing a deep learning-based speech enhancement model for real-world applications that meets the following criteria:
1. a small and fast model that can reduce single-frame real-time-factor (RTF) as much as possible while keeping competitive performance against the state-of-the-art deep learning networks, 2. a model that can perform both the denoising and derverberation simultaneously.

To address the first issue, we aim to improve a popular neural architecture, U-Net \cite{ronneberger2015u}, which has proven its superior performance on speech enhancement tasks \cite{choi2019phase,isik2020poconet, hu2020dccrn}.
The previous approaches that use U-Net on source separation applications apply convolution kernel not only on the frequency-axis but also on the time-axis.
This non-causal nature of U-Net increases computational complexity because additional computations are required on past and future frames to infer the current frame. 
Therefore, it is not suitable for online inference scenarios where the current frame needs to be processed in real-time.
In addition, the time-axis kernel makes the network computationally inefficient because there exists redundant computation between adjacent frames in both the encoding and decoding path of U-Net.
To tackle this problem, we propose a new neural architecture, Tiny Recurrent U-Net (TRU-Net), which is suitable for online speech enhancement.
The architecture is designed to enable efficient decoupling of the frequency-axis and time-axis computations, which makes the network fast enough to process a single frame in real-time. The number of parameters of the proposed network is only 0.38 million (M), which is small enough to deploy the model not only on a laptop but also on a mobile device and even on an embedded device combined with a quantization technique \cite{integeronlyquantization}.
The details of TRU-Net is described more in section \ref{sec:trunet}.

Next, to suppress the noise and reverberation simultaneously, we propose a phase-aware $\beta$-sigmoid mask (PHM).
The proposed PHM is inspired by \cite{wang2019deep}, in which the authors propose to estimate phase by reusing an estimated magnitude mask value from a trigonometric perspective.
The major difference between PHM and the approach in \cite{wang2019deep} is that PHM is designed to respect the triangular relationship between the mixture, the target source, and the remaining part, hence the sum of the estimated target source and the remaining part is always equal to the mixture.
We extend this property into a quadrilateral by producing two different PHMs simultaneously, which allows us to effectively deal with both denoising and dereverberation.
We will discuss PHM in further details in section \ref{sec:phm}.
\vspace{-0.1in}
\section{Tiny Recurrent U-Net}
\label{sec:trunet}

\vspace{-0.2in}
\begin{figure}[htbp]
\centering
\includegraphics[scale=0.4]{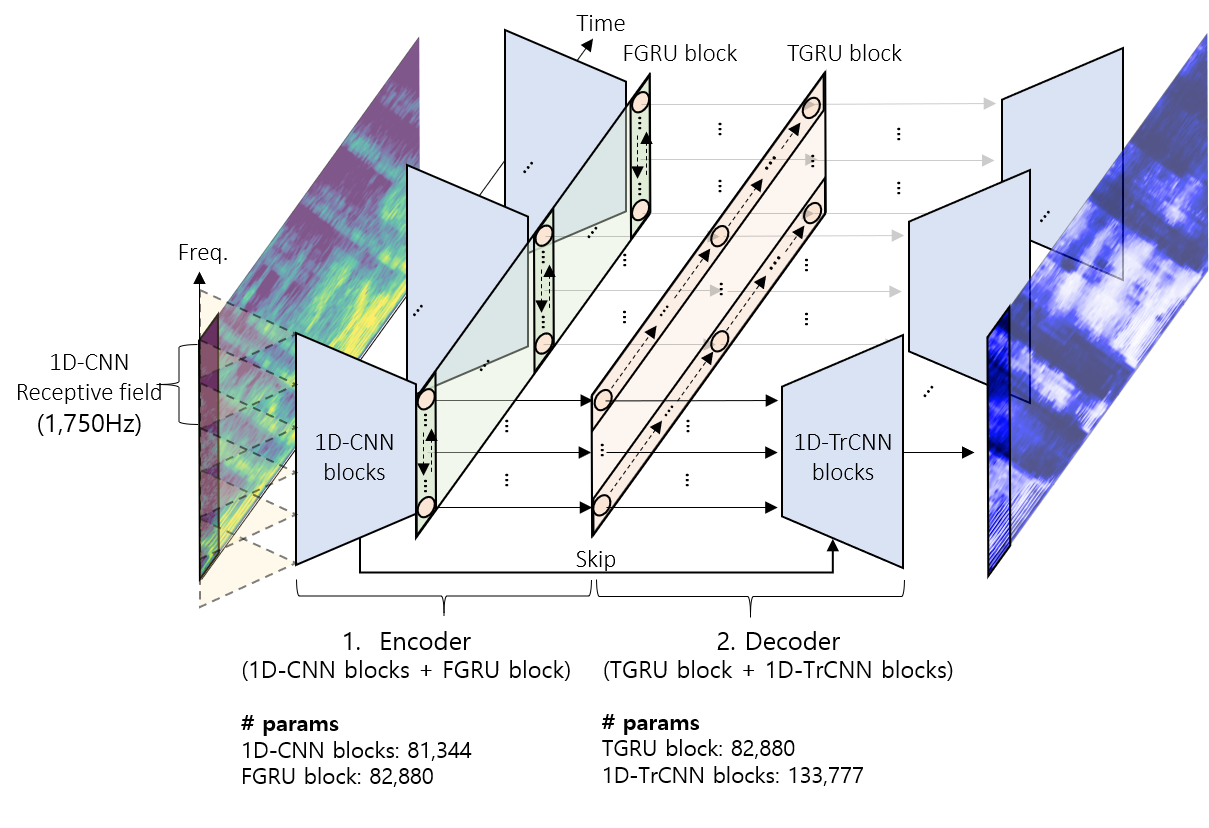}
\vspace{-0.2in}
\caption{The network architecture of TRU-Net}
\label{fig:trunet}
\end{figure}

\vspace{-0.25in}
\subsection{PCEN feature as an input}
\label{ssec:input}
A spectrogram is perhaps the most popular input feature for many speech enhancement models.
Per-channel energy normalization (PCEN) \cite{wang2017trainable} combines both dynamic range compression and automatic gain control together, which reduce the variance of foreground loudness and supress background noise when applied to a spectrogram \cite{lostanlen2018per}.
PCEN is also suitable for online inference scenarios as it includes a temporal integration step, which is essentially a first-order infinite impulse response filter that depends solely on a previous input frame.
In this work, we employ the trainable version of PCEN.

\vspace{-0.1in}
\subsection{Network architecture}
\label{ssec:architecture}
TRU-Net is based on U-Net architecture, except that the convolution kernel does not span the time-axis. Therefore, it can be considered a frequency-axis U-Net with 1D Convolutional Neural Networks (CNNs) and recurrent neural networks in the bottle-neck layer. 
The encoder is composed of 1D Convolutional Neural Network (1D-CNN) blocks and a Frequency-axis Gated Recurrent Unit (FGRU) block.
Each 1D-CNN block is a sequence of pointwise convolution and depthwise convolution similar to \cite{howard2017mobilenets}, except the first layer, which uses the standard convolution operation without a preceding pointwise convolution.
To spare the network size, we use six 1D-CNN blocks, which downsample the frequency-axis size from 256 to 16 using strided convolutions.
This results in a small receptive field (1,750Hz) which may be detrimental to the network performance.
To increase the receptive field, we use a bi-directional GRU layer \cite{cho-etal-2014-learning} along the frequency-axis instead of stacking more 1D-CNN blocks.
That is, the sequence of 16 vectors from 1D-CNN blocks is passed into the bi-directional GRU to increase the receptive field and share the information along the frequency-axis. We call this frequency-axis bi-directional GRU layer an FGRU layer.
A pointwise convolution, batch normalization (BN), and rectified linear unit (ReLU) are used after the FGRU layer, composing an FGRU block.
We used 64 hidden dimensions for each forward and backward FGRU cell.

The decoder is composed of a Time-axis Gated Recurrent Unit (TGRU) block and 1D Transposed Convolutional Neural Network (1D-TrCNN) blocks.
The output of the encoder is passed into a uni-directional GRU layer to aggregate the information along the time-axis. We call this GRU layer a TGRU layer.
While one can apply different GRU cells to each frequency-axis index of the encoder output, we shared the same cell on each frequency-axis index to save the number of parameters.
A pointwise convolution, BN, and ReLU follow the TGRU layer, composing a TGRU block.
We used 128 hidden dimensions for the TGRU cell.
Finally, 1D-TrCNN blocks are used to upsample the output from the TGRU block to the original spectrogram size.
The 1D-TrCNN block takes two inputs - 1. a previous layer output, 2. a skipped tensor from the encoder at the same hierarchy - and upsamples them as follows.
First, the two inputs are concatenated and projected to a smaller channel size (192 $\rightarrow$ 64) using a pointwise convolution.
Then, 1D transposed convolution is used to upsample the compressed information.
This procedure saves both the number of parameters and computation compared to the usual U-Net implementation where the two inputs are concatenated and upsampled immediately using the transposed convolution operation. Note that we did not use depthwise convolution for 1D-TrCNN block as we empirically observed that it drops the performance significantly when used in the decoding stage.

Every convolution operation used in the encoder and decoder is followed by BN and ReLU.
We denote the convolution configurations as follows, $l$-th: ($\kappa$, $s$, $c$)\,, where $l$, $\kappa$, $s$, $c$ denotes layer index, kernel size, strides, and output channels, respectively.
The detailed configurations of the encoder and decoder are as follows,
EncoderConfig = \{1-th: (5,2,64), 2-th: (3,1,128), 3-th: (5,2,128), 4-th: (3,1,128), 5-th: (5,2,128), 6-th: (3,2,128)\},
DecoderConfig = \{1-th: (3,2,64), 2-th: (5,2,64), 3-th: (3,1,64), 4-th: (5,2,64), 5-th: (3,1,64), 6-th: (5,2,10)\}.
Note that the pointwise convolution operations share the same output channel configuration with the exception that $\kappa$ and $s$ are both 1. The overview of TRU-Net and the number of parameters used for 1D-CNN blocks, FGRU block, TGRU block, and 1D-TrCNN blocks are shown in Fig. \ref{fig:trunet}.

\section{Single-stage Denoising and Dereverberation}
\label{sec:phm}
A noisy-reverberant mixture signal $\bm{x}$ is commonly modeled as the sum of additive noise $\bm{y}^{(n)}$ and reverberant source $\tilde{\bm{y}}$, where $\tilde{\bm{y}}$ is a result of convolution between room impulse response (RIR) $\bm{h}$ and dry source $\bm{y}$ as follows,
\begin{equation}
\bm{x} = \tilde{\bm{y}} + \bm{y}^{(n)} = \bm{h} \circledast \bm{y} + \bm{y}^{(n)}
\end{equation}
More concretely, we can break down $\bm{h}$ into two parts. First, the direct path part $\bm{h}^{(d)}$, which does not include the reflection path, and second, the rest of the part $\bm{h}^{(r)}$ including all the reflection paths as follows, 
\begin{equation}
\bm{x} = \bm{h}^{(d)} \circledast \bm{y} + \bm{h}^{(r)} \circledast \bm{y} + \bm{y}^{(n)} = \bm{y}^{(d)} + \bm{y}^{(r)} + \bm{y}^{(n)},    
\end{equation}
where $\bm{y}^{(d)}$ and $\bm{y}^{(r)}$ denotes a direct path source and reverberation, respectively.
In this setting, our goal is to separate $\bm{x}$ into three elements $\bm{y}^{(d)}$, $\bm{y}^{(r)}$, and $\bm{y}^{(n)}$.
Each of the corresponding time-frequency $(t,f)$ representations computed by short-time Fourier transform (STFT) is denoted as $X_{t,f} \in \mathbb{C}$, $Y^{(d)}_{t,f}\in \mathbb{C}$, $Y^{(r)}_{t,f} \in \mathbb{C}$, $Y^{(n)}_{t,f} \in \mathbb{C}$, and the estimated values will be denoted by the hat operator $\hat{\, \cdot \,}$. 

\vspace{-0.1in}
\subsection{Phase-aware \texorpdfstring{$\beta$}{}-sigmoid mask}
The proposed phase-aware $\beta$-sigmoid mask (PHM) is a complex-valued mask which is capable of systemically restricting the sum of estimated complex values to be exactly the value of mixture, $X_{t,f} = Y^{(k)}_{t,f} + Y^{(\lnot k)}_{t,f}$.
The PHM separates the mixture $X_{t,f}$ in STFT domain into two parts as \textit{one-vs-rest} approach, that is, the signal $Y^{(k)}_{t,f}$ and the sum of the rest of the signals $Y^{(\lnot k)}_{t,f} = X_{t,f}-Y^{(k)}_{t,f}$, where index $k$ could be one of the direct path source ($d$), reverberation ($r$), and noise ($n$) in our setting, $k \in \{d, r, n\}$.
The complex-valued mask $M^{(k)}_{t,f} \in \mathbb{C}$ estimates the magnitude and phase value of the source of interest $k$.

Computing PHM requires two steps. First, the network outputs the magnitude part of two masks $\lvert M^{(k)}_{t,f} \rvert$ and $\lvert M^{(\lnot k)}_{t,f} \rvert$ with sigmoid function $\sigma^{(k)}(\bm{z}_{t,f})$ multiplied by coefficient $\beta_{t,f}$ as follows, $\lvert M^{(k)}_{t,f} \rvert = \beta_{t,f} \cdot \sigma^{(k)}(\bm{z}_{t,f}) = \beta_{t,f} \cdot (1+e^{-(z^{(k)}_{t,f} - z^{(\lnot k)}_{t,f})})^{-1}$, 
where $z^{(k)}_{t,f}$ is the output located at $(t,f)$ from the last layer of neural-network function $\psi^{(k)}(\phi)$, and $\phi$ is a function composed of network layers before the last layer.
$\lvert M^{(k)}_{t,f} \rvert$ serves as a magnitude mask to estimate source $k$ and its value ranges from 0 to $\beta_{t,f}$.
The role of $\beta_{t,f}$ is to design a mask that is close to optimal values with a flexible magnitude range so that the values are not bounded between 0 and 1, unlike the commonly used sigmoid mask.
In addition, because the sum of the complex valued masks $M^{(k)}_{t,f}$ and $M^{(\lnot k)}_{t,f}$ must compose a triangle, it is reasonable to design a mask that satisfies the triangle inequalities, that is, $\lvert M^{(k)}_{t,f} \rvert + \lvert M^{(\lnot k)}_{t,f} \rvert$ $\geq 1$  and $\abs{ \lvert M^{(k)}_{t,f} \rvert - \lvert M^{(\lnot k)}_{t,f} \rvert} \leq 1$.
To address the first inequality we designed the network to output $\beta_{t,f}$ from the last layer with a softplus activation function as follows, $\beta_{t,f} = 1+ \texttt{softplus}((\psi_{\beta}(\phi))_{t,f})$, where $\psi_{\beta}$ denotes an additional network layer to output $\beta_{t,f}$. The second inequality can be satisfied by clipping the upper bound of the $\beta_{t,f}$ by $1 \mathbin{/} \lvert \, \sigma^{(k)}(\bm{z}_{t,f}) - \sigma^{(\lnot k)}(\bm{z}_{t,f}) \rvert$.

Once the magnitude masks are decided, we can construct a phase mask $e^{j\theta_{t,f}^{(k)}}$. 
Given the magnitudes as three sides of a triangle, we can compute the cosine of the absolute phase difference $\Delta \theta_{t,f}^{(k)}$ between the mixture and source $k$ as follows,
$\cos(\Delta \theta_{t,f}^{(k)}) = \nicefrac{ (1+{\lvert M_{t,f}^{(k)} \rvert}^2 - {\lvert M_{t,f}^{(\lnot k)}\rvert}^2 ) } {(2 \, \lvert M_{t,f}^{(k)}\rvert )}$.
Then, the rotational direction $\xi_{t,f} \in \{1, -1\}$ (clockwise or counterclockwise) for phase correction is estimated for the phase mask as follows, $e^{j\theta_{t,f}^{(k)}} = \cos(\Delta \theta_{t,f}^{(k)}) + j \xi_{t,f}\sin(\Delta \theta_{t,f}^{(k)})$.
Two-class straight-through Gumbel-softmax estimator was used to estimate $\xi_{t,f}$ \cite{DBLP:conf/iclr/JangGP17}. 
$M^{(k)}_{t,f}$ is defined as follows, $ M^{(k)}_{t,f} = \lvert M^{(k)}_{t,f} \rvert \cdot e^{j\theta_{t,f}^{(k)}}$.
Finally, $M^{(k)}_{t,f}$ is multiplied with $X_{t,f}$ to estimate the source $k$ as follows, $\hat{Y}^{(k)}_{t,f} = M^{(k)}_{t,f} \cdot X_{t,f}$.

\subsection{Masking from the perspective of a quadrilateral}
\begin{figure}[H]
\vspace{-0.1in}
\centering
\includegraphics[scale=0.4]{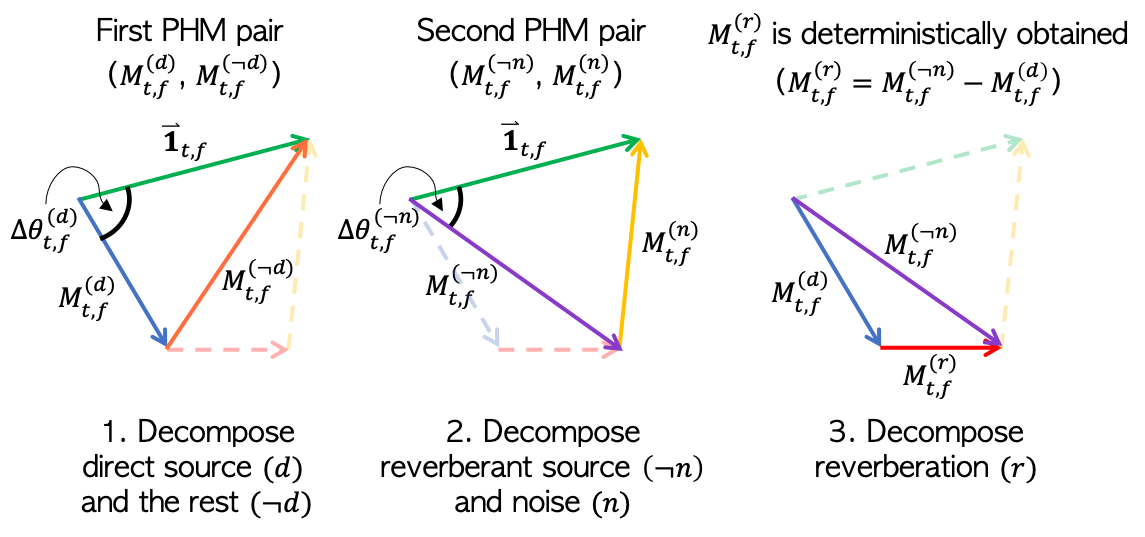}
\vspace{-0.1in}
\caption{The illustration of PHM masking method on a  quadrilateral}
\label{fig:quadrilateral}
\end{figure}
\vspace{-0.1in}
Because we desire to extract both the direct and reverberant source, two pairs of PHMs are used for each of them.
The first pair of masks, $M^{(d)}_{t,f}$ and $M^{(\lnot d)}_{t,f}$, separate the mixture into the direct source and the rest of the components, respectively.
The second pair of masks, $M^{(n)}_{t,f}$ and $M^{(\lnot n)}_{t,f}$, separate the mixture into the noise and the reverberant source.
Since PHM guarantees the mixture and separated components to construct a triangle in the complex STFT domain, the outcome of the separation can be seen from the perspective of a quadrilateral, as shown in Fig \ref{fig:quadrilateral}.
In this setting, as the three sides and two side angles are already determined by the two pairs of PHMs, the fourth side of the quadrilateral, $M^{(r)}_{t,f}$, is uniquely decided.

\vspace{-0.1in}
\subsection{Multi-scale objective}
Recently, a multi-scale spectrogram (MSS) loss function has been successfully used in a few audio synthesis studies \cite{wang2019neural, Engel2020DDSP}.
We incorporate this multi-scale scheme not only in the spectral domain but also in the waveform domain similar to \cite{Yao2019}.

Learning to maximize cosine similarity can be regarded as maximizing the signal-to-distortion ratio (SDR) \cite{choi2019phase}. Cosine similarity loss $C$ between the estimated signal $\hat{\bm{y}}^{(k)} \in \mathbb{R}^{N} $ and the ground truth signal $\bm{y}^{(k)} \in \mathbb{R}^{N}$  is defined as follows, $C(\bm{y}^{(k)},\hat{\bm{y}}^{(k)}) = -\frac{<\bm{y}^{(k)},\hat{\bm{y}}^{(k)}>}{\norm{\bm{y}^{(k)}}  \norm{\hat{\bm{y}}^{(k)}}}$, 
where $N$ denotes the temporal dimensionality of a signal and $k$ denotes the type of signal ($k \in \{d,r,n\}$).
Consider a sliced signal $\bm{y}^{(k)}_{[\frac{N}{M}(i-1):\frac{N}{M}i]}$, where $i$ denotes the segment index and $M$ denotes the number of segments.
By slicing the signal and normalizing it by its norm, each sliced segment is considered a unit for computing $C$.
 Therefore, we hypothesize that it is important to choose a proper segment length unit $\frac{N}{M}$ when computing $C$.
 In our case, we used multiple settings of segment lengths $g_{j}=\frac{N}{M_{j}}$ as follows,\useshortskip
\begin{equation}
\label{eq:multiscale}
	\mathcal{L}_{wav}^{(k)} = \sum_{j} \frac{1}{M_j}\sum_{i=1}^{M_j} C(\bm{y}^{(k)}_{[g_{j}(i-1):g_{j}i]},\hat{\bm{y}}^{(k)}_{[g_{j}(i-1):g_{j}i]}),
\end{equation}
\setlength{\belowdisplayskip}{0pt} \setlength{\belowdisplayshortskip}{0pt}
\setlength{\abovedisplayskip}{0pt} \setlength{\abovedisplayshortskip}{0pt}
where $M_{j}$ denotes the number of sliced segments. 
In our case, the set of $g_j$\textquotesingle s was chosen as follows, $g_j \in \{4064, 2032, 1016, 508\}$.

Next, the multi-scale loss on spectral domain is defined as follows,
\useshortskip
\begin{equation}
\mathcal{L}_{spec}^{(k)} = \sum_{i} \norm{ \, {\lvert  \text{STFT}_i(\bm{y}^{(k)}) \rvert} ^{0.3}-  {\lvert \text{STFT}_i(\hat{\bm{y}}^{(k)})}^{0.3} \rvert}^2,
\end{equation}
\useshortskip
where $i$ denotes the FFT size of $\text{STFT}_{i}$.
The only difference to the original MSS loss is that we replaced the log transformation into the power-law compression, as it has been successfully used in previous speech enhancement studies \cite{erdogan2018investigations, wilson2018exploring}.
We used the FFT sizes of STFT, (1024, 512, 256), with 75\% overlap.
The final loss function is defined by adding all the components as follows, $\mathcal{L}_\text{final} = \sum_{k \in \{d,r,n\}} \mathcal{L}_{wav}^{(k)} + \mathcal{L}_{spec}^{(k)}$.

\section{Experiments}
\label{sec:pagestyle}

\begin{table*}[t]
\begin{center}
\scalebox{0.68}{
    \centering
  \begin{tabular}{lcc|ccccccc|ccccccc} 
\toprule \\[-2ex]
$\,$ & $\,$ & $\,$ & \multicolumn{7}{c|}{Synthetic without Reverb} & \multicolumn{7}{c}{Synthetic with Reverb}\\
\midrule \\[-2ex]
Methods & Size(M/MB) & RT &{PESQ1} &{PESQ2} &{CBAK} &{COVL} &{CSIG} &{SI-SDR} &{STOI} &{PESQ1} &{PESQ2} &{CBAK}  &{COVL} &{CSIG} &{SI-SDR} &{STOI} \\
\midrule
Noisy & - & - & 2.45 & 1.58 & 2.53 & 2.35 & 3.19 & 9.07 & 91.52 & 2.75 & 1.82 & 2.80 & 2.64 & 3.50 & 9.03 & 86.62\\
NSnet \cite{xia2020weighted} & 1.27/4.84 & \cmark & 2.68 & 1.81 & 2.00 & 2.24 & 2.78 & 12.47 & 90.56 & 2.45 & 1.52 & 1.94 & 1.95 & 2.52 & 9.18 & 82.15 \\
DTLN \cite{westhausen2020dual} & 0.99/3.78 & \cmark & 3.04 & - & - & - & - & 16.34 & 94.76 & 2.70 & - & - & - & - & 10.53 & 84.68\\
ConvTasNet \cite{koyama2020exploring} & 5.08/19.38 & \xmark & - & 2.73 & 3.64 & 3.41 & 4.07 & - & - & - & 2.71 & \bf{3.67} & 3.47 & \bf{4.21} & - & - \\
PoCoNet1 \cite{isik2020poconet} & 50/190.73 & \xmark & - & 2.71 & 3.02 & 3.29 & 3.85 & - & - & - & \bf{2.83} & 3.21 & 3.35 & 3.83 & - & - \\
PoCoNet2 \cite{isik2020poconet} & 50/190.73 & \xmark & - & 2.75 & 3.04 & 3.42 & 4.08 & - & - & - & - & - & - & - & - & - \\
DCCRN-E \cite{hu2020dccrn} & 3.7/14.11 & \cmark & 3.27 & - & - & - & - & - & - &  3.08 & - & - & - & - & - & -\\
DCCRN-CL \cite{hu2020dccrn} & 3.7/14.11 & \xmark & 3.26 & - & - & - & - & - & - & 3.10 & - & - & - & - & - & -\\
\midrule
\bf{TRU-Net} (FP32) & 0.38/1.45 & \cmark & \bf{3.36} & \bf{2.86} & \bf{3.66} & \bf{3.55} & \bf{4.21} & \bf{17.55} & \bf{96.32} &  \bf{3.35} & 2.74 & 3.62 & \bf{3.48} & 4.17 & \bf{14.87} & \bf{91.29}\\
\bf{TRU-Net} (INT8) & 0.38/0.36 & \cmark & 3.35 & 2.84 & 3.62 & 3.53 & 4.18 & 17.23 & 96.12 & 3.31 & 2.70 & 3.56 & 3.45 & 4.16 & 14.47 & 91.01\\
\bottomrule
    \end{tabular}
}
\vspace{-0.1in}
\caption{Objective evaluation results on DNS-challenge synthetic development sets. PoCoNet2 denotes the model with partial dereverberation described in \cite{isik2020poconet}, and PoCoNet1 is the model trained without it. We denote the network size (Size) in two aspects, the number of parameters in million (M) and the actual model size in megabyte (MB). The models with real-time (RT) capability are marked with \cmark, otherwise \xmark.
}
\label{tab:dns}
\vspace{-0.2in}
\end{center}
\end{table*}

\subsection{Implementation details}
Since our goal is to perform both denoising and dereverberation, we used pyroomacoustics \cite{scheibler2018pyroomacoustics} to simulate an artificial reverberation with randomly sampled absorption, room size, location of source and microphone distance.
We used 2 seconds of speech and noise segments, and mixed them with a uniformly distributed source-to-noise ratio (SNR) ranging from -5 dB to 25 dB.
Input features were used as a channel-wise concatenation of log-magnitude spectrogram, PCEN spectrogram, and real/imaginary part of demodulated phase.
We used AdamW optimizer \cite{DBLP:conf/iclr/ReddiKK18} and the learning rate was halved when the validation score did not improve for three consecutive epochs. The initial learning rate was set to 0.0004.
The window size and hop size were set to 512 (32 ms) and 128 (8 ms), respectively.

	We also quantized the proposed model into INT8 format and compared the model size with prior works. The purpose of our quantized model experiments is to reduce the model size and computational cost for embedded environments. We adopted the computation flow using quantized numbers suggested in \cite{integeronlyquantization} to quantize the neural network. In addition, the uniform symmetric quantization scheme \cite{googlewhitepaper}, which uses uniform quantization and restricts zero-point to 0, was applied for efficient hardware implementation. In the experiments, all the layers in the neural network are processed using quantized weights, activations, and inputs; only bias values are represented in full precision. Other processing steps such as feature extraction and masking are computed in full precision. For encoder and decoder layers, we observe the scale statistics of intermediate tensors during training. Then, during inference, we fix the scales of activations using the average of the observed minimum and maximum values. Only GRU layers are dynamically quantized during the inference time due to the large dynamic range of internal activations at each time step.

\vspace{-0.1in}
\subsection{Ablation study}
\label{ssec:ablation study}
In order to confirm the effect of PCEN, multi-scale objective, and FGRU block, we trained and validated the model using the CHiME2 training set and development set, respectively. 
An ablation study was conducted on the CHiME2 test set.
TRU-Net-A denotes the proposed method. TRU-Net-B denotes the model trained without multi-scale objective. TRU-Net-C denotes the model trained without the PCEN feature. TRU-Net-D denotes the model trained without FGRU block.
We used the original SDR \cite{vincent2006performance} to compare our model with other models.
The results are shown in Table \ref{tab:chime2}. 
It is clearly observable that all the proposed methods are contributing to performance improvement. Note that FGRU block contributes significantly on the performance. 
We also compared the proposed model with other models using the CHiME2 test set.
The proposed model showed better performance than not only the recent lightweight model Tiny-LSTM (TLSTM) and its pruned version (PTLSTM) \cite{fedorov2020tinylstms}, but also the large-sized model \cite{wilson2018exploring}.

\begin{table}[htbp]
\renewcommand{\tabcolsep}{1.6mm}
\begin{center}
\scalebox{0.68}{
\centering
\begin{tabular}{lc|ccccccc} 
\toprule \\[-2ex]
$\,$ & $\,$ & \multicolumn{7}{c}{Input SNR}\\
\midrule \\[-2ex]
Methods & Size (M/MB) & {-6} &{-3} &{0} &{3} &{6} &{9} & Avg.\\
\midrule
TLSTM (FP32) \cite{fedorov2020tinylstms} & 0.97/3.70 & 10.01  & 11.54 & 13.08 & 14.23 & 15.85 & 17.46 & 13.70 \\
PTLSTM (FP32) \cite{fedorov2020tinylstms} & 0.52/1.97 &  10.07  & 11.59 & 13.10 & 14.31 &  15.89  & 17.50 & 13.74 \\
PTLSTM (INT8) \cite{fedorov2020tinylstms} & 0.61/0.58 &  9.82  & 11.37 & 12.91 & 14.20 & 15.74 & 17.44 &  13.58\\
PTLSTM (INT8) \cite{fedorov2020tinylstms} & 0.33/0.31 &  9.33  &  10.91 & 12.46 & 13.79 & 15.46 & 17.16  & 13.18 \\
Wilson et al. \cite{wilson2018exploring} & 65/247.96 & 12.17  & 13.44  & 14.70 & 15.83 & 17.30 & 18.78 & 15.37 \\
\midrule
TRU-Net-A (FP32) & 0.38/1.45 & \bf{12.36} & \bf{13.62} & \bf{15.08} & \bf{16.21} & \bf{17.70} & \bf{19.39} & \bf{15.73}\\
TRU-Net-B (FP32) & 0.38/1.45 & 12.21 & 13.39 & 14.91 & 16.09 & 17.53 & 19.24 & 15.56\\
TRU-Net-C (FP32) & 0.38/1.45 & 11.96 & 13.24 & 14.69 & 15.97 & 17.47 & 19.18 & 15.42\\
TRU-Net-D (FP32) & 0.31/1.18 & 11.83 & 13.14 & 14.63 & 15.85 & 17.28 & 18.97 & 15.28\\
TRU-Net-A (INT8) & 0.38/0.36 & 12.35 & 13.62 & 15.03 & 16.18 & 17.62 & 19.30 & 15.68\\
TRU-Net-B (INT8) & 0.38/0.36 & 12.23 & 13.40 & 14.91 & 16.08 & 17.51 & 19.21 & 15.56\\
TRU-Net-C (INT8) & 0.38/0.36 & 11.96 & 13.20 & 14.64 & 15.94 & 17.42 & 19.11 & 15.38\\
TRU-Net-D (INT8) & 0.31/0.30 & 11.79 & 13.13 & 14.56 & 15.78 & 17.19 & 18.85 & 15.22\\
\bottomrule
    \end{tabular}
}
\vspace{-0.1in}
\caption{Objective evaluation results on the CHiME2 test set.}
\label{tab:chime2}
\vspace{-0.4in}
\end{center}
\end{table}

\vspace{-0.1in}
\subsection{Denoising results}
\label{ssec:denoising_results}
We further checked the denoising performance of our model by training the model on the large scale DNS-challenge dataset \cite{reddy2020icassp} and internally collected dataset.
It was tested on two non-blind DNS development sets, 1) synthetic clips without reverb (Synthetic without Reverb) and 2) synthetic clips with reverb (Synthetic with Reverb).
We compared our model with the recent models \cite{isik2020poconet, hu2020dccrn, xia2020weighted, westhausen2020dual, koyama2020exploring} submitted to the previous 2020 Interspeech DNS-challenge.
6 evaluation metrics, PESQ, CBAK, COVL, CSIG, SI-SDR, and STOI \cite{recommendation2001perceptual, loizou2013speech, le2019sdr, taal2010short}, were used.
Note that although it is recommended to use ITU-T P862.2 wide-band version of PESQ (PESQ2), a few studies reported their score using ITU-T P862.1 (PESQ1).
Therefore, we used both PESQ versions to compare our model with other models.
The results are shown in Table \ref{tab:dns}.
We can see that TRU-Net shows the best performance in the Synthetic without Reverb set while having the smallest number of parameters.
In the Synthetic with Reverb set, TRU-Net showed competitive performance using orders of magnitude fewer parameters than other models.

\vspace{-0.1in}
\subsection{Dereverberation results}

\label{ssec:dereverb_results}
The performance of simultaneous denoising and dereverberation was tested on \textit{min} subset of WHAMR dataset, which contains 3,000 audio files.
The WHAMR dataset is composed of noisy-reverberant mixtures and the direct sources as ground truth.
TRU-Net models (FP32 and INT8) in Table \ref{tab:dns} were used for the test.
We show the denoising and dereverberation performance of our model in Table \ref{tab:whamr} along with two other models that were tested on the same WHAMR dataset.
Our model achieved the best results compared to the other baseline models, which shows the parameter efficiency of TRU-Net on simultaneous denoising and dereverberation task.

\vspace{-0.1in}
\begin{table}[htbp]
\begin{center}
\scalebox{0.68}{
    \centering
  \begin{tabular}{l|c|ccc}
    \toprule
Method & Size (M/MB) & PESQ1 & SI-SDR & STOI \\
\midrule
Noisy & - & 1.83  & -2.73  & 73.00  \\
NSnet \cite{xia2020weighted} & 1.27/4.84 & 1.91  & 0.34  & 73.02  \\
DTLN \cite{westhausen2020dual} & 0.99/3.78 & 2.23  & 2.12  & 80.40  \\
\midrule
TRU-Net (FP32) & 0.38/1.45 & \bf{2.51} & \bf{3.51} & \bf{81.22} \\
TRU-Net (INT8) & 0.38/0.36 & 2.49 & 3.03 & 80.56 \\
\bottomrule
    \end{tabular}
}
\vspace{-0.1in}
\caption{
Objective evaluation of simultaneous denoising and dereverberation results on the WHAMR dataset.
}
\label{tab:whamr}
\end{center}
\vspace{-0.1in}
\end{table}

\vspace{-0.2in}
\subsection{Listening test results}
\label{ssec:listening}
Using the proposed model (TRU-Net (FP32)) in Table \ref{tab:dns}, we participated in 2021 ICASSP DNS Challenge Track 1 \cite{reddy2020icassp}.
For better perceptual quality, we mixed the estimated direct source and reverberant source at 15 dB, and applied a zero-delay dynamic range compression (DRC).
The average computation time to process a single frame (including FFT, iFFT, and DRC) took 1.97 ms and 1.3 ms on 2.7 GHz Intel i5-5257U and 2.6 GHz Intel i7-6700HQ CPUs, respectively.
The lookahead of TRU-Net is 0 ms.
The listening test was conducted based on ITU-T P.808.
The results are shown in Table \ref{tab:p808}.
The model was tested on various speech sets including singing voice, tonal language, non-English (includes tonal), English, and emotional speech. The results show that TRU-Net can achieve better performance than the baseline model, NSnet2 \cite{braun2020data}.
\begin{table}[htbp]
\renewcommand{\tabcolsep}{1.6mm}
\begin{center}
\scalebox{0.68}{
    \centering
  \begin{tabular}{lc|cccccc}
    \toprule
Method & Size (M/MB) & Singing & Tonal & Non-English & English & Emotional & Overall \\
\midrule
Noisy & - & 2.96 & 3.00 & 2.96 & 2.80 & 2.67 & 2.86 \\
NSnet2 \cite{braun2020data} & 2.8/10.68 & \bf{3.10} & 3.25 & 3.28 & 3.30 & \bf{2.88} & 3.21\\
\midrule
TRU-Net & 0.38/1.45 & 3.08 & \bf{3.38} & \bf{3.43} & \bf{3.41} & \bf{2.88} & \bf{3.32} \\
\bottomrule
    \end{tabular}
}
\vspace{-0.1in}
\caption{
MOS results on the DNS-challenge blind test set
}
\label{tab:p808}
\end{center}
\vspace{-0.1in}
\end{table}

\vspace{-0.3in}
\section{Relation to prior works}
\vspace{-0.1in}
\label{sec:related_work}
Recently, there has been increasing interest in phase-aware speech enhancement because of the sub-optimality of reusing the phase of the mixture signal.
While most of these works tried to estimate the clean phase by using a phase mask or an additional network, the absolute phase difference between mixture and source can be actually computed using the law of cosines \cite{mowlaee2012phase}.
Inspired by this, \cite{wang2019deep} proposed to estimate a rotational direction of the absolute phase difference for speech separation. 

The FGRU and TGRU used in TRU-Net is similar to the work in \cite{grzywalski2019using}. They used bidirectional long short-term memory (bi-LSTM) networks on the frequency-axis and the time-axis combined with 2D-CNN-based U-Net. 
The difference is that bi-LSTM was utilized to increase performance in \cite{grzywalski2019using}, whereas we employ FGRU and uni-directional TGRU to better handle the online inference scenario combined with the proposed lightweight 1D-CNN-based (frequency-axis) U-Net.
\vspace{-0.1in}
\section{Conclusions}
\vspace{-0.1in}
\label{sec:print}
In this work, we proposed TRU-Net, which is an efficient neural network architecture specifically designed for online inference applications.
Combined with the proposed PHM, we successfully demonstrated a single-stage denoising and dereverberation in real-time.
We also showed that using PCEN and multi-scale objectives improves the performance further.
Experimental results confirm that our model achieve comparable performance with state-of-the-art models having a significantly larger number of parameters.
For future work, we plan to employ modern pruning techniques on an over-parameterized model to develop a big-sparse model which may provide better performance than a small-dense model with the same number of parameters.

\vfill\pagebreak
\bibliographystyle{IEEEbib}
\bibliography{strings,refs}

\begin{thebibliography}{10}

\bibitem{ronneberger2015u}
Olaf Ronneberger, Philipp Fischer, and Thomas Brox,
\newblock ``U-net: Convolutional networks for biomedical image segmentation,''
\newblock in {\em Proc. MICCAI}, 2015, pp. 234--241.

\bibitem{choi2019phase}
Hyeong-Seok Choi, Jang-Hyun Kim, Jaesung Huh, Adrian Kim, Jung-Woo Ha, and
  Kyogu Lee,
\newblock ``Phase-aware speech enhancement with deep complex u-net,''
\newblock {\em arXiv preprint arXiv:1903.03107}, 2019.

\bibitem{isik2020poconet}
Umut Isik, Ritwik Giri, Neerad Phansalkar, Jean-Marc Valin, Karim Helwani, and
  Arvindh Krishnaswamy,
\newblock ``Poconet: Better speech enhancement with frequency-positional
  embeddings, semi-supervised conversational data, and biased loss,''
\newblock in {\em Proc. INTERSPEECH}, 2020.

\bibitem{hu2020dccrn}
Yanxin Hu, Yun Liu, Shubo Lv, Mengtao Xing, Shimin Zhang, Yihui Fu, Jian Wu,
  Bihong Zhang, and Lei Xie,
\newblock ``Dccrn: Deep complex convolution recurrent network for phase-aware
  speech enhancement,''
\newblock in {\em Proc. INTERSPEECH}, 2020.

\bibitem{integeronlyquantization}
Benoit Jacob, Skirmantas Kligys, Bo~Chen, Menglong Zhu, Matthew Tang, Andrew
  Howard, Hartwig Adam, and Dmitry Kalenichenko,
\newblock ``Quantization and training of neural networks for efficient
  integer-arithmetic-only inference,''
\newblock in {\em Proc. CVPR}, 2018, pp. 2704--2713.

\bibitem{wang2019deep}
Zhong-Qiu Wang, Ke~Tan, and DeLiang Wang,
\newblock ``Deep learning based phase reconstruction for speaker separation: A
  trigonometric perspective,''
\newblock in {\em Proc. ICASSP}, 2019, pp. 71--75.

\bibitem{wang2017trainable}
Yuxuan Wang, Pascal Getreuer, Thad Hughes, Richard~F Lyon, and Rif~A Saurous,
\newblock ``Trainable frontend for robust and far-field keyword spotting,''
\newblock in {\em Proc. ICASSP}, 2017, pp. 5670--5674.

\bibitem{lostanlen2018per}
Vincent Lostanlen, Justin Salamon, Mark Cartwright, Brian McFee, Andrew
  Farnsworth, Steve Kelling, and Juan~Pablo Bello,
\newblock ``Per-channel energy normalization: Why and how,''
\newblock {\em IEEE Signal Processing Letters}, vol. 26, no. 1, pp. 39--43,
  2018.

\bibitem{howard2017mobilenets}
Andrew~G Howard, Menglong Zhu, Bo~Chen, Dmitry Kalenichenko, Weijun Wang,
  Tobias Weyand, Marco Andreetto, and Hartwig Adam,
\newblock ``Mobilenets: Efficient convolutional neural networks for mobile
  vision applications,''
\newblock {\em arXiv preprint arXiv:1704.04861}, 2017.

\bibitem{cho-etal-2014-learning}
Kyunghyun Cho, Bart van Merri{\"e}nboer, Caglar Gulcehre, Dzmitry Bahdanau,
  Fethi Bougares, Holger Schwenk, and Yoshuas Bengio,
\newblock ``Learning phrase representations using {RNN} encoder{--}decoder for
  statistical machine translation,''
\newblock in {\em Proc. EMNLP}, 2014, pp. 1724--1734.

\bibitem{DBLP:conf/iclr/JangGP17}
Eric Jang, Shixiang Gu, and Ben Poole,
\newblock ``Categorical reparameterization with gumbel-softmax,''
\newblock in {\em Proc. ICLR}, 2017.

\bibitem{wang2019neural}
Xin Wang, Shinji Takaki, and Junichi Yamagishi,
\newblock ``Neural source-filter-based waveform model for statistical
  parametric speech synthesis,''
\newblock in {\em Proc. ICASSP}, 2019, pp. 5916--5920.

\bibitem{Engel2020DDSP}
Jesse Engel, Lamtharn~(Hanoi) Hantrakul, Chenjie Gu, and Adam Roberts,
\newblock ``Ddsp: Differentiable digital signal processing,''
\newblock in {\em Proc. ICLR}, 2020.

\bibitem{Yao2019}
Jian Yao and Ahmad Al-Dahle,
\newblock ``{Coarse-to-Fine Optimization for Speech Enhancement},''
\newblock in {\em Proc. INTERSPEECH}, 2019, pp. 2743--2747.

\bibitem{erdogan2018investigations}
Hakan Erdogan and Takuya Yoshioka,
\newblock ``Investigations on data augmentation and loss functions for deep
  learning based speech-background separation.,''
\newblock in {\em INTERSPEECH}, 2018, pp. 3499--3503.

\bibitem{wilson2018exploring}
Kevin Wilson, Michael Chinen, Jeremy Thorpe, Brian Patton, John Hershey, Rif~A
  Saurous, Jan Skoglund, and Richard~F Lyon,
\newblock ``Exploring tradeoffs in models for low-latency speech enhancement,''
\newblock in {\em IWAENC}, 2018, pp. 366--370.

\bibitem{xia2020weighted}
Yangyang Xia, Sebastian Braun, Chandan~KA Reddy, Harishchandra Dubey, Ross
  Cutler, and Ivan Tashev,
\newblock ``Weighted speech distortion losses for neural-network-based
  real-time speech enhancement,''
\newblock in {\em Proc. ICASSP}, 2020, pp. 871--875.

\bibitem{westhausen2020dual}
Nils~L Westhausen and Bernd~T Meyer,
\newblock ``Dual-signal transformation lstm network for real-time noise
  suppression,''
\newblock in {\em Proc. INTERSPEECH}, 2020.

\bibitem{koyama2020exploring}
Yuichiro Koyama, Tyler Vuong, Stefan Uhlich, and Bhiksha Raj,
\newblock ``Exploring the best loss function for dnn-based low-latency speech
  enhancement with temporal convolutional networks,''
\newblock {\em arXiv preprint arXiv:2005.11611}, 2020.

\bibitem{scheibler2018pyroomacoustics}
Robin Scheibler, Eric Bezzam, and Ivan Dokmani{\'c},
\newblock ``Pyroomacoustics: A python package for audio room simulation and
  array processing algorithms,''
\newblock in {\em Proc. ICASSP}, 2018, pp. 351--355.

\bibitem{DBLP:conf/iclr/ReddiKK18}
Sashank~J. Reddi, Satyen Kale, and Sanjiv Kumar,
\newblock ``On the convergence of adam and beyond,''
\newblock in {\em Proc. ICLR}, 2018.

\bibitem{googlewhitepaper}
Raghuraman Krishnamoorthi,
\newblock ``Quantizing deep convolutional networks for efficient inference: A
  whitepaper,''
\newblock {\em arXiv preprint arXiv:1806.08342}, 2018.

\bibitem{vincent2006performance}
Emmanuel Vincent, R{\'e}mi Gribonval, and C{\'e}dric F{\'e}votte,
\newblock ``Performance measurement in blind audio source separation,''
\newblock {\em IEEE transactions on audio, speech, and language processing},
  vol. 14, no. 4, pp. 1462--1469, 2006.

\bibitem{fedorov2020tinylstms}
Igor Fedorov, Marko Stamenovic, Carl Jensen, Li-Chia Yang, Ari Mandell, Yiming
  Gan, Matthew Mattina, and Paul~N Whatmough,
\newblock ``Tinylstms: Efficient neural speech enhancement for hearing aids,''
\newblock in {\em Proc. INTERSPEECH}, 2020.

\bibitem{reddy2020icassp}
Chandan~KA Reddy, Harishchandra Dubey, Vishak Gopal, Ross Cutler, Sebastian
  Braun, Hannes Gamper, Robert Aichner, and Sriram Srinivasan,
\newblock ``Icassp 2021 deep noise suppression challenge,''
\newblock {\em arXiv preprint arXiv:2009.06122}, 2020.

\bibitem{recommendation2001perceptual}
ITU-T Recommendation,
\newblock ``Perceptual evaluation of speech quality (pesq): An objective method
  for end-to-end speech quality assessment of narrow-band telephone networks
  and speech codecs,''
\newblock {\em Rec. ITU-T P. 862}, 2001.

\bibitem{loizou2013speech}
Philipos~C Loizou,
\newblock {\em Speech enhancement: theory and practice},
\newblock CRC press, 2013.

\bibitem{le2019sdr}
Jonathan Le~Roux, Scott Wisdom, Hakan Erdogan, and John~R Hershey,
\newblock ``Sdr--half-baked or well done?,''
\newblock in {\em Proc. ICASSP}, 2019, pp. 626--630.

\bibitem{taal2010short}
Cees~H Taal, Richard~C Hendriks, Richard Heusdens, and Jesper Jensen,
\newblock ``A short-time objective intelligibility measure for time-frequency
  weighted noisy speech,''
\newblock in {\em Proc. ICASSP}, 2010, pp. 4214--4217.

\bibitem{braun2020data}
Sebastian Braun and Ivan Tashev,
\newblock ``Data augmentation and loss normalization for deep noise
  suppression,''
\newblock in {\em International Conference on Speech and Computer}, 2020, pp.
  79--86.

\bibitem{mowlaee2012phase}
Pejman Mowlaee, Rahim Saeidi, and Rainer Martin,
\newblock ``Phase estimation for signal reconstruction in single-channel source
  separation,''
\newblock in {\em Thirteenth Annual Conference of the International Speech
  Communication Association}, 2012.

\bibitem{grzywalski2019using}
Tomasz Grzywalski and Szymon Drgas,
\newblock ``Using recurrences in time and frequency within u-net architecture
  for speech enhancement,''
\newblock in {\em Proc. ICASSP}, 2019, pp. 6970--6974.

\end{thebibliography}

\end{document}